\documentclass[aps,prl,twocolumn,superscriptaddress,showpacs]{revtex4}
\usepackage{graphicx}
\usepackage{graphicx,epstopdf,color}
\usepackage{amsfonts}
\usepackage{amsmath,amssymb,mathrsfs}
\usepackage{bm}
\usepackage{float}
\usepackage{pst-grad}
\usepackage{dcolumn}
\pacs{74.70.Xa, 72.15.-v}

\begin{document}

\title{Effect of disorder on the resistivity anisotropy near the electronic nematic phase transition in pure and electron-doped BaFe$_2$As$_2$}

\affiliation{Geballe Laboratory for Advanced Materials and Department of Materials Science and Engineering, Stanford University, USA}
\affiliation{Geballe Laboratory for Advanced Materials and Department of Applied Physics, Stanford University, USA}
\affiliation{Stanford Institute for Materials and Energy Sciences, SLAC National Accelerator Laboratory,\\ 2575 Sand Hill Road, Menlo Park, CA 94025, USA}

\author{Hsueh-Hui Kuo}
\affiliation{Geballe Laboratory for Advanced Materials and Department of Materials Science and Engineering, Stanford University, USA}
\affiliation{Stanford Institute for Materials and Energy Sciences, SLAC National Accelerator Laboratory,\\ 2575 Sand Hill Road, Menlo Park, CA 94025, USA}

\author{Ian R. Fisher}
\affiliation{Geballe Laboratory for Advanced Materials and Department of Applied Physics, Stanford University, USA}
\affiliation{Stanford Institute for Materials and Energy Sciences, SLAC National Accelerator Laboratory,\\ 2575 Sand Hill Road, Menlo Park, CA 94025, USA}

\begin{abstract}

We show that the strain-induced resistivity anisotropy in the tetragonal state of the representative underdoped Fe-arsenides BaFe$_2$As$_2$, Ba(Fe$_{1-x}$Co$_x$)$_2$As$_2$ and Ba(Fe$_{1-x}$Ni$_x$)$_2$As$_2$ is independent of disorder over a wide range of defect and impurity concentrations. This result demonstrates that the anisotropy in the in-plane resistivity in the paramagnetic orthorhombic state of this material is not due to elastic scattering from anisotropic defects. Conversely, our result can be most easily understood if the resistivity anisotropy arises primarily from an intrinsic anisotropy in the electronic structure.

\end{abstract}

\maketitle

Ongoing experimental investigations reveal that the underdoped regime of the cuprate high-temperature superconductors harbors a variety of poorly understood broken symmetry states. In the case of the ferro-pnictide and chalcogenide superconductors, the broken symmetries are much clearer \cite{Kivelson_2008}, but the physical origin of the phase transitions is still a subject of debate \cite{CCLee_2009, Kruger_2009, Bascones_2010, Yin_2010, Lv_2010, CCChen_2010, Laad_2011, Mazin_2009, Fang_2008, Xu_2008, Fernandes_2010, Fernandes_2011, Fernandes_2012, Fernandes_Schmalian_2012, Daghofer_2010, Dagotto_2013, Lee_2009}. Of particular interest, the ferropnictides suffer a tetragonal-to-orthorhombic structural transition at a temperature $T_s$ that either precedes or accompanies the onset of long range antiferromagnetic magnetic order at $T_N$ (see Ref. \cite{Ian_review} and references therein). From the perspective of symmetry, all physical properties develop a two-fold in-plane anisotropy at such a phase transition. However, the magnitude depends on microscopic details, and therefore measurements that probe the anisotropy in the broken symmetry state can directly or indirectly inform our understanding of the mechanism that drives the phase transition. Quantities such as the in-plane resistivity anisotropy are therefore of considerable interest, and it is especially important to establish intrinsic versus extrinsic effects. 

In this paper, we show for several representative underdoped Fe-pnictides that the strain-induced resistivity anisotropy in the tetragonal state is independent of the degree of disorder for a given value of $T_N$ over a wide range of defect and impurity concentrations. This result can be directly compared to the anisotropy that develops spontaneously in the orthorhombic state (Appendix II)\cite{Ian_review, JH_2010, Tanatar_2010, Davis_2010, Liang_2011}, and therefore demonstrates that the in-plane resistivity anisotropy observed for this family of compounds in the paramagnetic orthorhombic state is not an extrinsic effect associated with defect scattering. The result can be most easily understood if the resistivity anisotropy in this regime is primarily determined by the Fermi surface anisotropy rather than an anisotropy in the scattering rate.

The structural phase transition that occurs in underdoped Fe-pnictides breaks a point symmetry of the original crystal lattice, and hence free-standing crystals naturally form structural twins in order to minimize the elastic energy \cite{Ian_review}. The in-plane anisotropy can nevertheless be probed using uniaxial stress to detwin single crystals, as has now been done for several different families \cite{Ian_review, JH_2010, Tanatar_2010, Davis_2010, Blomberg_2011, FeTe_detwin}. These measurements reveal two principal results. First, the resistivity anisotropy for some compositions can rise to very large values; for example, $\rho_b/ \rho_a \sim 2$ for as-grown crystals of Ba(Fe$_{0.965}$Co$_{0.035}$)$_2$As$_2$ \cite{JH_2010}, even though the degree of orthorhombicity is relatively modest ($(a-b)/[(1/2)(a+b)] \sim 0.35 \%$ for the specific case cited \cite{Prozorov_2009}). Second, the materials exhibit a large stress-induced anisotropy in the tetragonal state \cite{JH_2012, HH_2013, Blomberg_2012}- an effect that we will return to shortly. That said, two important results call in to question whether these effects are truly intrinsic to the orthorhombic state and/or symmetry. Specifically, measurements of annealed crystals of Ba(Fe$_{1-x}$Co$_x$)$_2$As$_2$ held under uniaxial stress indicate that the resistivity anisotropy diminishes after annealing \cite{Liang_2011, Nakajima_2012, Ishida_2013}. The experiments did not check the degree of detwinning, nor was the stress or strain measured, so quantitative comparisons are unfortunately impossible, but nevertheless this observation suggests that elastic scattering might be significant in determining the resistivity anisotropy. Furthermore, STM measurements reveal extended anisotropic defects at low temperature, perhaps associated with impurities that locally polarize the electronic structure \cite{Davis_2010, Davis_2013}. Both of these observations suggest that the resistivity anisotropy might be a parasitic effect associated with anisotropic elastic scattering from extended defects. However, electronic reconstruction below $T_N$ results in small Fermi surface pockets with a very different character to the original bands \cite{Suchitra_QO, Jim_QO, Terashima_2009}. Consequently, the resistivity anisotropy deep in the antiferromagnetic state is not the ideal quantity to inform discussion of the physical origin of the structural phase transition that occurs at a much higher temperature. 

Ideally, one would measure the resistivity anisotropy in the temperature window between $T_s$ and $T_N$. However, quantitative analysis relies on knowledge of the exact degree of detwinning, which is difficult to monitor for the entire volume of material that is probed by the electrical resistivity. Furthermore, the range of temperature between $T_N$ and $T_s$ is small, in many cases just a few K. To avoid these difficulties, we probe the induced resistivity anisotropy that occurs in the tetragonal state as a consequence of anisotropic biaxial strain. This not only reveals the electronic anisotropy associated with orthorhombic symmetry in the absence of twin domains and magnetic order, but also probes the important fluctuation regime above $T_s$. We do this first for the undoped parent compound BaFe$_2$As$_2$, comparing measurements for samples with different residual resistivity values, and second for Co and Ni-substituted samples, comparing samples for which the structural, magnetic and superconducting transitions coincide, but for which the impurity scattering rate is very different.  

\begin{figure}
\includegraphics[width=8cm]{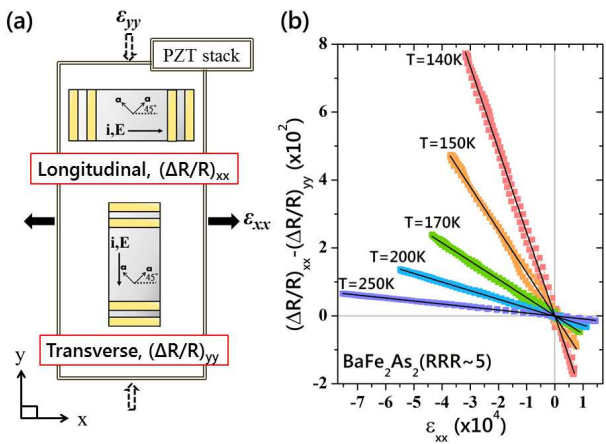} 
\caption{(Color online) (a) Schematic diagram illustrating measurement of longitudinal elastoresistance, $(\Delta R/R)_{xx}$ (i.e. current $\parallel$ $\epsilon_{xx}$) and transverse elastoresistance $(\Delta R/R)_{yy}$ (i.e. current $ \perp$ $\epsilon_{xx}$ ) for the case of $\epsilon_{xx}$ aligned along the [110]$_T$ tetragonal crystallographic direction. Gold rectangles indicate position of electrical contacts for standard four-point measurement. Actual crystal dimensions are typically $0.5 \times 0.1 mm$, compared with the PZT stack which has lateral dimensions $9.15 \times 5.2 mm$ (b) Representative data showing the induced resistivity anisotropy ($N \sim (\Delta R/R)_{xx}-(\Delta R/R)_{yy}$) as a function of strain $\epsilon_{xx}$ at several temperatures above $T_s$ for BaFe$_2$As$_2$ with RRR $\sim$ 5. Black lines show linear fits for each temperature from which the elatoresistivity coefficient $m_{66}$ is extracted.} 
\end{figure} 

Our experiments are based on measurements of the elastoresistivity coefficients of the materials in question. The elastoresistance of a solid describes changes in the electrical resistance as a consequence of the strains experienced by the solid. In linear response, appropriate for the small strains developed in this experiment, the relative change in resistivity is given by $(\Delta\rho/\rho)_i = \displaystyle\sum_{j=1}^{6} m_{ij}\epsilon_j$, where $xx = 1, yy = 2, zz = 3, yz = 4, zy = 5, xy = 6$. For tetragonal symmetry, the elastoresistivity tensor $m_{ij}$ has six independent coefficients, which can be determined through a combination of elastoresistance measurements using different sample orientations \cite{HH_2013}. We use commercially available PZT piezoelectric stacks (PSt150/5x5/7 cryo 1 from Piezomechanik) to generate anisotropic biaxial in-plane strain, following the general method described in Ref.s \cite{HH_2013}, \cite{JH_2012}, and \cite{Butkovicova_glue}. Crystals are glued to the surface of the piezoelectric stack using five minute epoxy (from ITW Devcon), and electrical contact is made using silver paste onto evaporated gold contacts for standard four-point resistance measurements. For the specific experimental geometry employed ($\epsilon_{xx} \parallel$ [110]$_T$) (indicated in Fig. 1(a)) the induced resistivity anisotropy that occurs due to the anisotropic biaxial strain is given by $N =(\rho_a-\rho_b)/[(1/2)(\rho_a+\rho_b)] \sim (\Delta R/R)_{xx}-(\Delta R/R)_{yy} = (1+\nu_p)2m_{66}\epsilon_{xx}$ \cite{HH_2013}. Hence, simultaneous measurements of the longitudinal $(\Delta R/R)_{xx}$ and transverse elastoresistance $(\Delta R/R)_{yy}$ directly yields the induced resistivity anisotropy, which is proportional to the elastoresistivity coefficient $m_{66}$.  The strain $\epsilon_{xx}$ is measured by a strain gauge glued on the back surface of the PZT stack, and separate measurements of larger crystals were used to ensure that the strain was fully transmitted through the samples \cite{JH_2012}. The Poisson's ratio of the piezoelectric stack, $\nu_p$, is almost temperature independent and was characterized by separate measurements of mutually transverse strain gauges \cite{HH_2013}.   

Single crystals of BaFe$_2$As$_2$ were grown from a ternary flux as described previously \cite{Mandrus_2008, JH_2009}. As-grown crystals have a residual resistance ratio RRR = R(300K)/R(0K) $\sim$ 3 - 5, and a structural/N\'{e}el transition temperature T$_{s/N} \sim$ 134 K. Crystals were also annealed in vacuum at 700$^\circ$C for four weeks, resulting in a substantial decrease in the residual resistivity (and corresponding increase in RRR to a value of $\sim$ 15.5, illustrated in Fig. 2(a)) and a modest change in the structural/N\'{e}el transition $T_{s/N} \sim$ 138 K. Representative data showing the induced resistivity anisotropy $N$ as a function of strain for as-grown BaFe$_2$As$_2$ are shown in Fig. 1(b). The slope of the linear fit of $N$ vs $\epsilon_{xx}$ at each temperature yields $(1+\nu_p)2m_{66}$ from which the temperature dependence of the elastoresistivity coefficient $2m_{66}$ can be readily extracted using measured values of $\nu_p$. Experiments were repeated for several samples and representative data are shown in Fig. 2(b) for as-grown and annealed samples. For temperatures greater than $T_{s/N}$, the elastoresistivity coefficients (i.e. the induced anisotropy) for as-grown and annealed samples are identical within experimental resolution (see also fit parameters in Appendix I). For temperatures below $T_{s/N}$ extrinsic effects associated with twin domain motion dominate the elastoresistance coefficients, which are different for the two RRR values, perhaps reflecting differences in pinning. The following discussion focuses on the behavior for temperatures greater than $T_s$, for which the measured elastoresistivity is determined solely by the intrinsic properties of the tetragonal structure.   

\begin{figure}
\includegraphics[width=8cm]{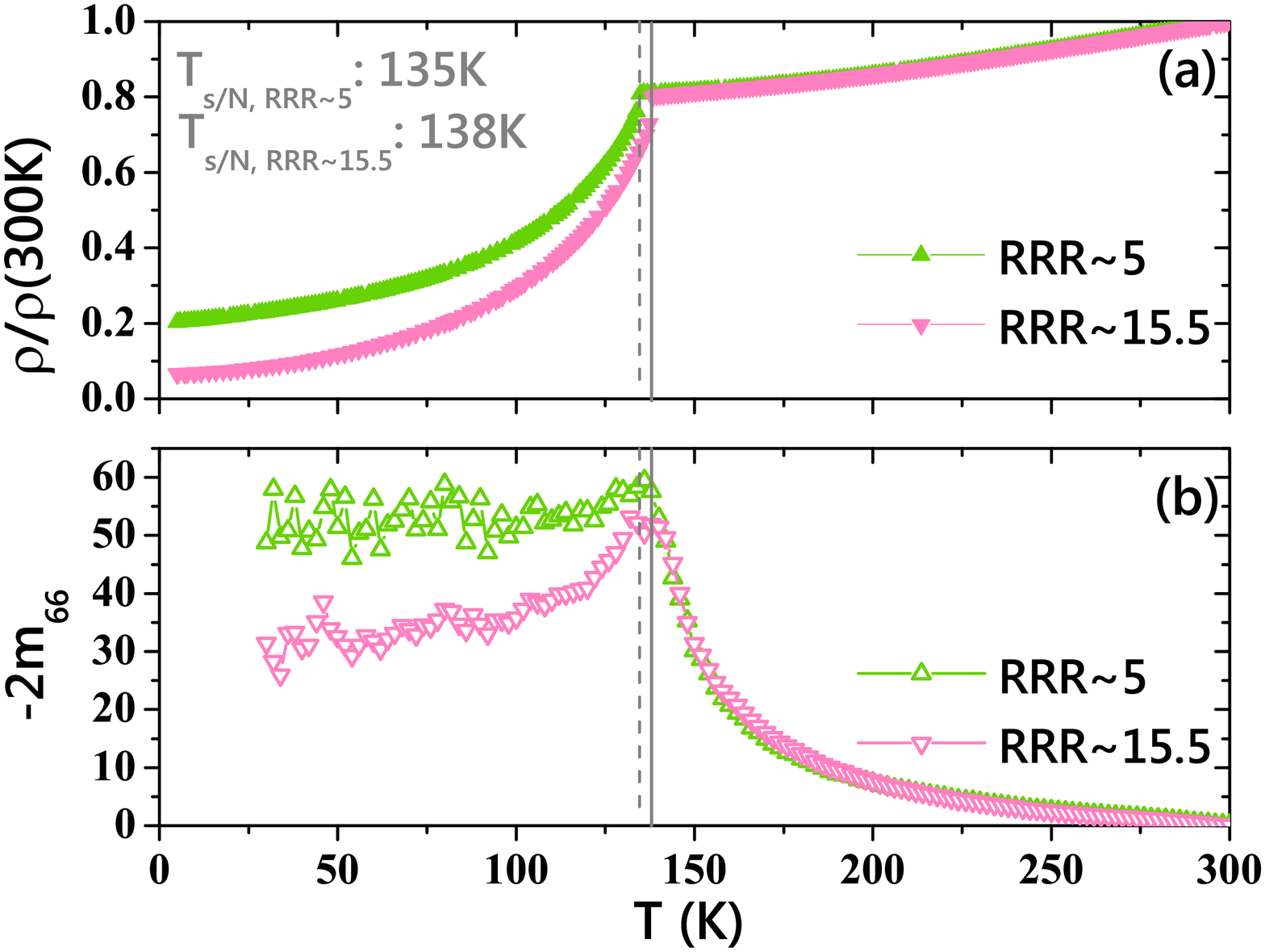} 
\caption{(Color online) (a) Temperature dependence of the normalized resistivity, $\rho/\rho(300K)$ for BaFe$_2$As$_2$ with RRR $\sim$ 5 (as grown samples, shown by green data points) and RRR $\sim$ 15.5 (annealed samples, pink data points). (b) Temperature-dependence of the elastoresistivity coefficients $-2m_{66}$ for the same samples shown in (a). $T_{s,N}$ values determined from $d\rho/dT$ are indicated by dashed and solid vertical lines for as-grown and annealed samples respectively.} 
\end{figure}

\begin{figure}
\includegraphics[width=8cm]{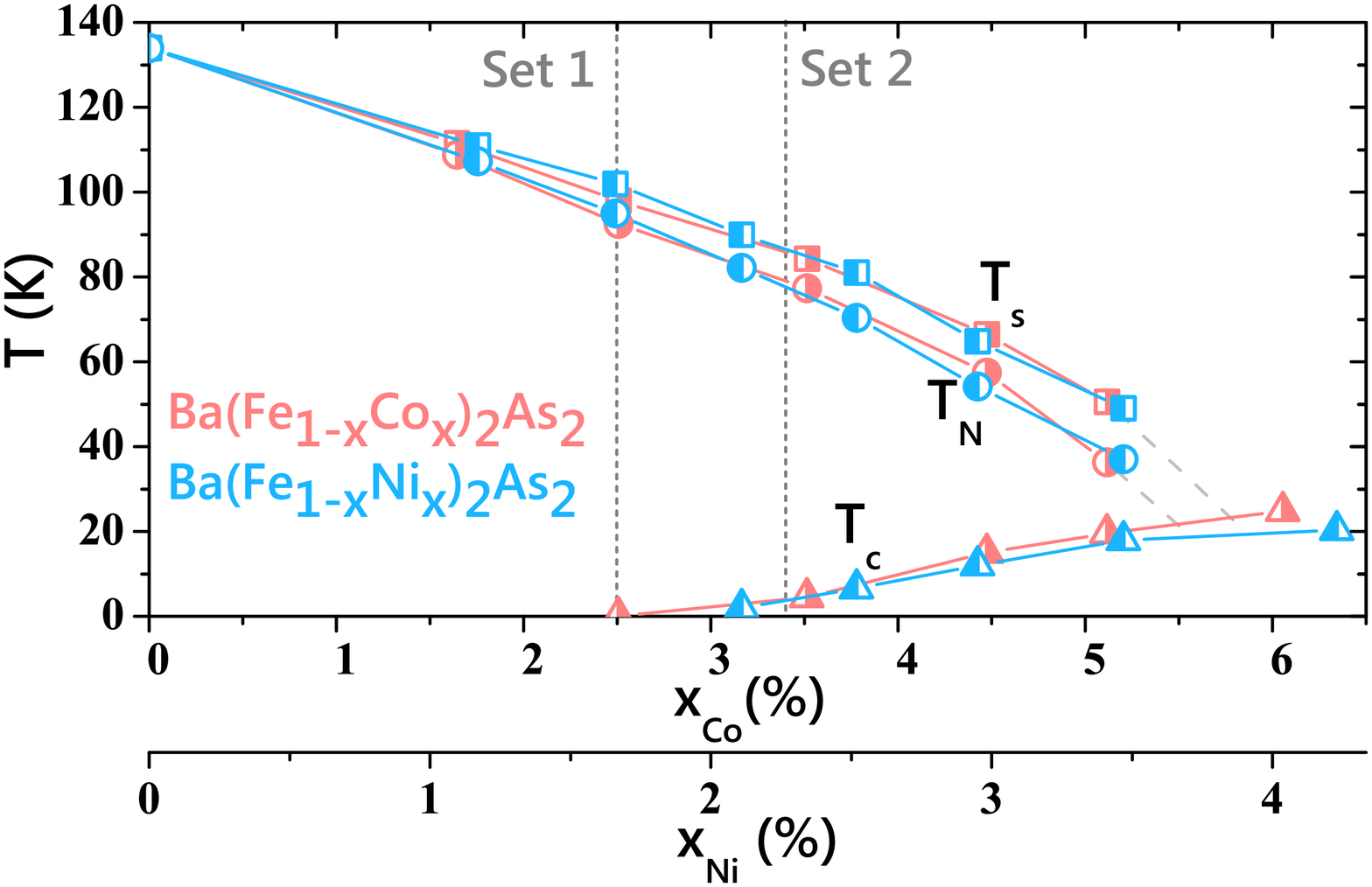} 
\caption{(Color online) Phase diagrams of Ba(Fe$_{1-x}$Co$_x$)$_2$As$_2$ (red) and Ba(Fe$_{1-x}$Ni$_x$)$_2$As$_2$ (blue). Squares, circles and triangles indicate T$_s$, T$_N$, and T$_c$ respectively. T$_s$ and T$_N$ were determined from $d\rho/dT$ of free standing samples. $T_c$ was defined by the midpoint of the superconducting transitions. The two sets of compositions demonstrated in this paper (2.5$\%$ Co, 1.7$\%$ Ni, and 3.4$\%$ Co, 2.1$\%$ Ni) are denoted by dotted lines.} 
\end{figure}

\begin{figure*}
\includegraphics[width=8cm]{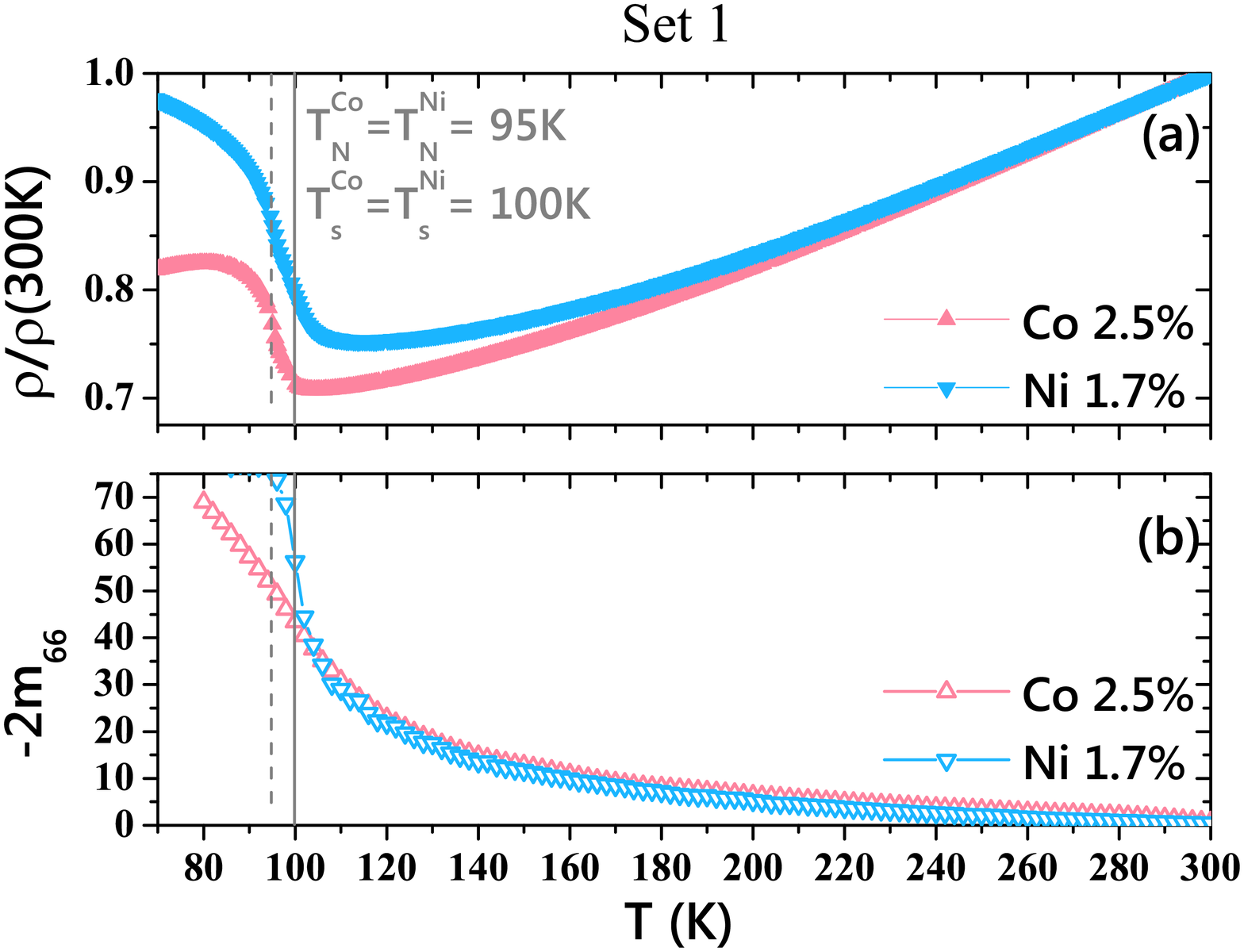}
\qquad
\includegraphics[width=8cm]{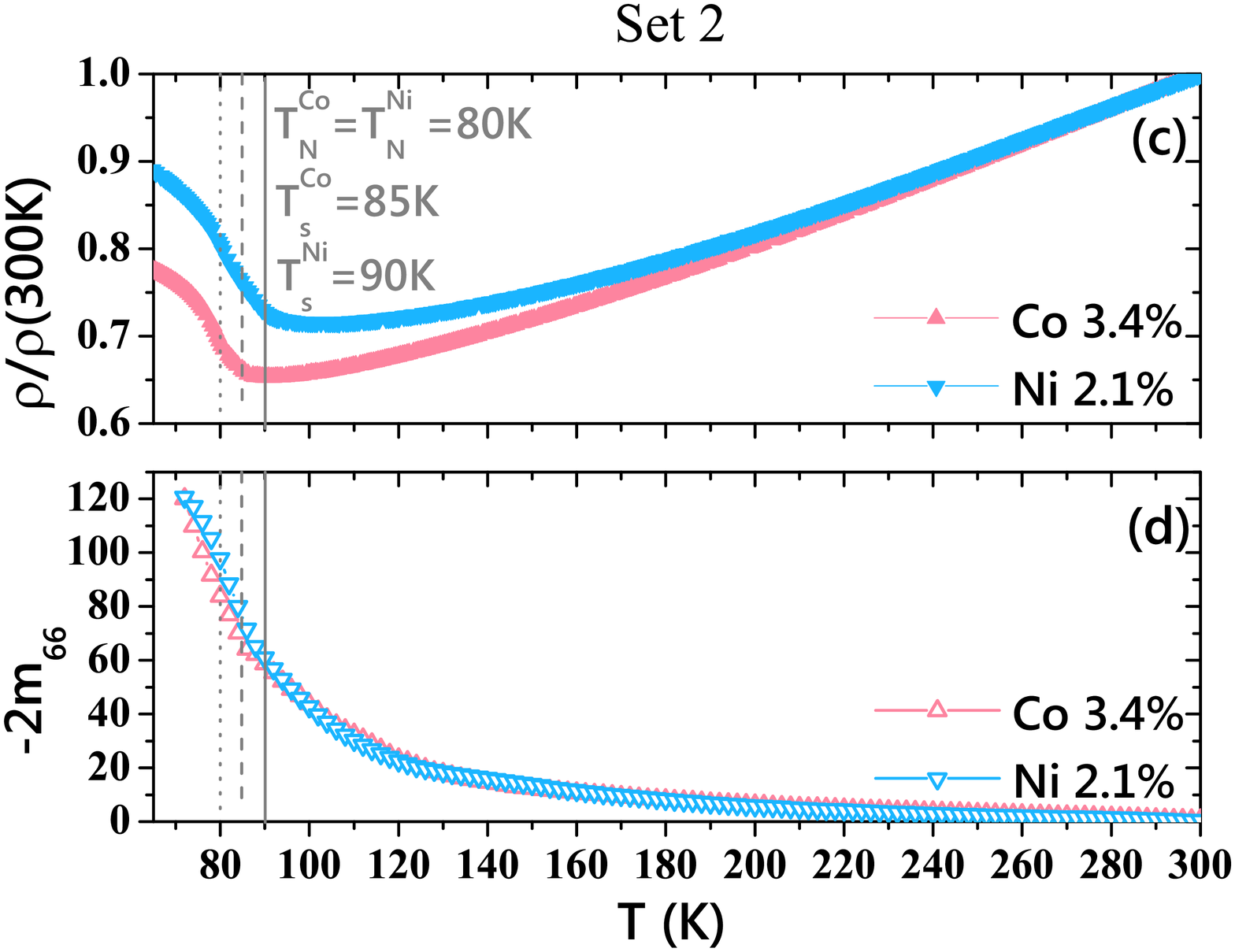} 
\caption{(Color online) Comparison of elastoresistivity coefficients for Co and Ni doped samples with identical $T_N$ values. (a,b) Temperature dependence of the normalized resistance, $\rho/\rho(300K)$ and the elastoresistivity coefficient $-2m_{66}$ respectively for samples from Set 1 ($T_N$ = 95 K). (c,d) Similar data data for samples from Set 2 ($T_N$ = 80 K). Solid and dashed vertical lines indicate $T_s$ and $T_N$ respectively for each composition. For both sets of samples, $m_{66}$ is identical for the Co and Ni doped samples, despite the significant difference in impurity scattering.} 
\end{figure*} 

The apparent insensitivity to disorder revealed in Fig. 2(b) motivated us to explore the effects of stronger impurity scattering associated with chemical substitution. We chose cobalt and nickel substitution as two cases that are well characterized and for which the depth of the impurity potential is rather different. Band structure calculations reveal that both impurities effectively increase the Fermi level (i.e. electron-dope), with nickel having a larger effect due to the increased number of electrons per impurity \cite{Ideta_2013}. Empirically, the phase diagrams of Ba(Fe$_{1-x}$Co$_x$)$_2$As$_2$ and Ba(Fe$_{1-x}$Ni$_x$)$_2$As$_2$ are found to be very similar \cite{HH_2011}, and for these underdoped compositions the composition-dependence of the critical temperatures associated with the separated structural, N{\'e}el and superconducting transitions ($T_s$, $T_N$ and $T_c$ respectively) scale almost exactly if the dopant concentration is scaled by a factor of 1.6 (i.e. $T_{i}^{Co}$ $\sim$ $T_{i}^{Ni}$ if $x_{Co}$ $\sim$ 1.6 $x_{Ni}$, as shown in Fig. 3). In order to separate effects due to disorder from changes in the band filling, we specifically compare $m_{66}$ coefficients of Co and Ni-doped samples for which the critical temperatures coincide, implying that their electronic structures are comparable. However, as we will show, our main result does not rely on this assumption. We chose two sets of compositions, comparing Ba(Fe$_{0.975}$Co$_{0.025}$)$_2$As$_2$ with Ba(Fe$_{0.983}$Ni$_{0.017}$)$_2$As$_2$ (both of which have $T_N$= 95K, labeled Set 1), and comparing Ba(Fe$_{0.966}$Co$_{0.034}$)$_2$As$_2$ with Ba(Fe$_{0.979}$Ni$_{0.021}$)$_2$As$_2$ (both with $T_N$=80K, labeled Set 2). For Set 1 ($T_N$ = 95 K), $T_s$ is identical for both Co and Ni doped samples. For Set 2, $T_s$ differs by 5 K for the same value of $T_N$ ($T_N$ = 80 K). 
 
Single crystals of Ba(Fe$_{1-x}$Co$_x$)$_2$As$_2$ and Ba(Fe$_{1-x}$Ni$_x$)$_2$As$_2$ were grown from a pseudo-ternary melt, similar to the parent compound and following established protocols \cite{Mandrus_2008, JH_2009}. The composition was determined by electron microprobe analysis, with an uncertainty in the dopant concentration of 0.0015. Critical temperatures were determined from the derivative of the resistivity \cite{JH_2010}, and are shown by vertical lines in Fig. 4 for Sets 1 and 2. We normalize values of the resistivity at 300K to avoid uncertainty due to geometric factors. For both sets of samples, the resistivity of the Ni-doped samples was larger than that of the Co-doped samples for the same values of $T_N$ (panels (a) and (c) of Fig. 4) even though the absolute dopant concentration was lower, consistent with the deeper impurity potential associated with Ni impurities relative to Co. In particular, the RRR of the Ni-doped samples is smaller than that of the Co-doped samples with the same $T_N$, as has previously been observed (Appendix I and \cite{HH_2011}). In other words, for a given value of $T_N$, impurity scattering is stronger for Ni doped samples relative to Co-doped samples \cite{Olariu_2011}.

Following the same procedure as for the parent compound, we also measured the induced resistivity anisotropy, described by the elastoresistivity coefficient $2m_{66}$ as a function of temperature. Representative data are shown for the two sets of compositions in panels (b) and (d) of Fig. 4. Similar to the parent compound, we find that for $T > T_s$ the elastoresistivity coefficient is identical within the resolution of the measurement (fit parameters given in Appendix I). Apparently the induced anisotropy in the tetragonal state is independent of disorder for a given value of $T_N$, at least over the range of compositions studied here. This is our main result.

To understand the significance of the above results, we first review the origin of the striking temperature-dependence of the $m_{66}$ elastoresistivity coefficients seen in Fig.s 2, and 4. As was shown recently, $m_{66}$ for these heavily underdoped compositions diverges following a Curie-Weiss temperature-dependence ($2m_{66} = \lambda/[a_0(T-T^*)] + 2m_{66}^0$) over a wide range of temperatures, spanning from $T_s$ up to room temperature \cite{HH_2013, JH_2012}. This unusual result (for ordinary metals the elastoresistivity coefficients are small and essentially temperature independent \cite{Kuczynski}) is a direct consequence of the presence of an electronic nematic order parameter $\psi$, which aligns in the strain field. For sufficiently small values of the nematic order parameter the resistivity anisotropy is linearly proportional (i.e. $N \propto \psi$; this follows from symmetry, since $N$ changes sign if the $a$ and $b$ axes are reversed in the orthorhombic state \cite{resistor_network} and has also been shown explicitly via a Boltzmann transport analysis \cite{Fernandes_2011}). Furthermore, since shear strain ($\gamma_{ab}$ relative to the tetragonal axes, equivalent to $\epsilon_{xx} - \epsilon_{yy}$ in the experimental coordinate system) couples linearly to the nematic order parameter, the component of the nematic susceptibility tensor relevant for
spontaneous nematic order in the [110]$_T$ direction of the tetragonal system (i.e. the B$_{2g}$ channel) is given by $\cfrac{\partial \psi}{\partial \gamma_{ab}}\propto \cfrac{\partial N}{\partial\gamma_{ab}} = 2m_{66}$ \cite{HH_2013}. In other words, the elastoresistivity coefficient $2m_{66}$ directly measures the nematic susceptibility ($\chi_N$) for the B$_{2g}$ shear channel up to a multiplicative constant ($2m_{66} = c\chi_N$). The eventual pseudo-proper ferroelastic phase transition is driven by the growing nematic fluctuations, revealed by the divergence of the nematic susceptibility, which in mean field follows a Curie-Weiss temperature dependence \cite{HH_2013}.
 
Our experiments reveal that the induced resistivity anisotropy associated with anisotropic in-plane biaxial strain (i.e. orthorhombicity (see Appendix II)), which is given by $m_{66}$ and is proportional to the nematic susceptibility, is independent of disorder for a given value of $T_N$. The nematic susceptibility and the proportionality constant (c) that relates it to $m_{66}$ could, at least in principle, depend upon disorder. However, it is difficult to conceive of a physical mechanism by which their mutual effect would be to leave their product (i.e. 2$m_{66}$) unaffected by disorder for all three cases considered ($T_N \sim$ 134K, 95K and 80K) unless each quantity were itself independent of disorder, at least over the range of disorder considered here. Our results therefore strongly constrain any models of the resistivity anisotropy in this material.

Impurity scattering has been invoked in different contexts to account for the in-plane resistivity anisotropy in the Fe pnictides. At low temperatures, the apparent reduction in the anisotropy following annealing treatments \cite{Liang_2011, Nakajima_2012} points towards an important role for impurity scattering in the N{\'e}el state, perhaps due to extended anisotropic defects \cite{Davis_2010, Davis_2013}. However, our result shows that in the paramagnetic orthorhombic state (which is
arguably the more important regime for addressing questions related to nematic order), the resistivity anisotropy is independent of disorder. We therefore conclude that arguments based on anisotropic elastic scattering have at best only limited validity, restricted to the low-temperature N{\'e}el state. Impurity scattering can, however, affect the resistivity anisotropy in other, more subtle ways. In models based on spin-fluctuation scattering, quenched disorder can affect the relative contributions to the conductivity from hot spots and cold regions of the Fermi surface, indirectly affecting the resistivity anisotropy in the nematic phase. These ideas have been used to  predict a reversal of the sign of the resistivity anisotropy for hole doped cases \cite{Fernandes_2011, Fernandes_Schmalian_2012}, which was recently observed experimentally \cite{Blomberg_2013}. As mentioned above, for this scenario to be operative would require fine tuning such that disorder-induced changes in the product of $\chi_N$ and the proportionality constant c relating it to the resistivity anisotropy exactly balance each other for all three cases considered, which is somewhat unsatisfying \cite{constant}. In contrast, our result can be readily understood if the resistivity anisotropy is primarily determined by the Fermi surface anisotropy, itself directly related to the nematic order parameter.

In conclusion, we stress our main experimental finding, which is that for a given value of $T_N$ the strain-induced resistivity anisotropy in the tetragonal state is independent of disorder for the representative underdoped Fe-arsenides BaFe$_2$As$_2$, Ba(Fe$_{1-x}$Co$_x$)$_2$As$_2$ and Ba(Fe$_{1-x}$Ni$_x$)$_2$As$_2$. The resistivity anisotropy in the paramagnetic orthorhombic state is therefore an intrinsic property of the material, and consequently any succesful theory must account for the large resistivity anisotropy that has been observed for detwinned samples in this regime. 

\section{Acknowledgments}

The authors thank S. A. Kivelson, R. M. Fernandes, A. Moreo and E. Dagotto for helpful conversations. This work was supported by the DOE, Office of Basic Energy Sciences, under Contract No. DE-AC02-76SF00515.

\appendix

\section{Appendix I: Fit parameters}

\section{Fit parameters}

As discussed in the main text, the divergence of the elastoresistivity coefficient, $m_{66}$, above $T_s$ follows a Curie-Weiss temperature-dependence ($2m_{66} = \lambda/[a_0(T-T^*)] + 2m_{66}^0$). Table 1 lists the fit parameters for each composition shown in the main text (Fig. 2: BaFe$_2$As$_2$, RRR$\sim$5 and BaFe$_2$As$_2$, RRR$\sim$15.5; Fig. 4, Set 1:   Ba(Fe$_{0.975}$Co$_{0.025}$)$_2$As$_2$ and Ba(Fe$_{0.983}$Ni$_{0.017}$)$_2$As$_2$; Fig. 4, Set 2: Ba(Fe$_{0.966}$Co$_{0.034}$)$_2$As$_2$ and Ba(Fe$_{0.979}$Ni$_{0.021}$)$_2$As$_2$). For each set of samples with the same $T_N$, neither the magnitude of the resistivity anisotropy (i.e. $\lambda / a_0)$ nor the bare mean-field nematic transition temperature $T^*$, show an effect of disorder within experimental uncertainty.

RRR values for Co and Ni doped samples from Set 1 and 2 are 1.32 and 0.98, and 1.25 and 1.02, respectively. (Here we approximate the RRR using the resistivity at room temperature and 25K due to the onset of superconductivity.) In other words, for a given value of $T_N$, the RRR is lower for the Ni-doped samples than the Co-doped samples. In both cases we have used as-grown crystals. 

\begin{table}[h]
\centering
\begin{tabular*}{0.465\textwidth}{@{\extracolsep{\fill}} l|ccc}
Sample & 2m$_{66}^0$ & $\lambda /a_0 (K)$ & T$^*(K)$ \\ \hline \hline
BaFe$_2$As$_2$ &&&\\ 
RRR $\sim$ 5 & 4.7 $\pm$ 0.5 & -897 $\pm$ 43 & 124.9 $\pm$ 0.8 \\ 
RRR $\sim$ 15.5 & 4.9 $\pm$ 0.3 & -942 $\pm$ 32 & 124.8 $\pm$ 0.6 \\ \hline 
Set 1 &&&\\ 
Co 2.5$\%$ & 3.6 $\pm$ 0.3 & -1238 $\pm$ 46 & 73.7 $\pm$ 1 \\ 
Ni 1.7$\%$  & 4.5 $\pm$ 0.6 & -1221 $\pm$ 90 & 73.6 $\pm$ 2.3 \\ \hline
Set 2 &&&\\ 
Co 3.4$\%$ & 5.6 $\pm$ 1.2 & -1528 $\pm$ 162 & 68.1 $\pm$ 2.6 \\ 
Ni 2.1$\%$  & 5.6 $\pm$ 0.8 & -1529 $\pm$ 103 & 67.8 $\pm$ 1.7 \\ \hline
\end{tabular*}
\caption{Fit parameters from the fit of $2m_{66}$ with $2m_{66} = \lambda/[a_0(T-T^*)] + 2m_{66}^0$ for all the compositions shown in Fig. 2 and Fig. 4.}
\end{table}

\section{Shear strain and orthorhombicity}

At the structural phase transition, the material develops a spontaneous anisotropic in-plane strain relative to the isotropic tetragonal state, characterized by a finite difference of the in-plane lattice constants $a$ and $b$ (where $a$ and $b$ refer to the orthorhombic lattice). In contrast, our experiments induce a finite anisotropic biaxial in-plane strain in the tetragonal state along [110]$_T$, such that $a = a_0(1 + \epsilon_{xx})$ and $b = b_0(1 + \epsilon_{yy}) = b_0(1 - \nu_p\epsilon_{xx})$, where $a_0=b_0$ in the tetragonal state and $\nu_p$ is the Poisson's ratio of the piezoelectric stack. The induced orthorhombicity is therefore $O = \frac{a-b}{(1/2)(a+b)} \sim (\epsilon_{xx}-\epsilon_{yy}) = (1+\nu_p)\epsilon_{xx}$. Measured relative to the original tetragonal crystal axes, which are oriented at 45$^o$ to the orthorhombic axes, this is equal to the engineering strain $\gamma_{ab}$. (For a discussion of the definition of the engineering sheer strain, see for example, J. F. Nye "Physcial properties of crystals", (Oxford University Press, 1972).) For typical values of $\epsilon_{xx}$ ($\sim 3\times10^{-4}$) this yields $O \sim 10^{-3}$,  comparable to the spontaneous strain that develops at $T_s$. 

Symmetry dictates that in the orthorhombic phase, the resistivity anisotropy $N =(\rho_a-\rho_b)/[(1/2)(\rho_a+\rho_b)]$ is linearly proportional to the orthorhombicity $O$ for small values of the order parameter, (this is experimentally verified \cite{Blomberg_2012}).  Similarly, in the tetragonal state, the induced resistivity anisotropy is linearly proportional to the anisotropic biaxial strain; $N \propto (\epsilon_{xx}-\epsilon_{yy})$ (i.e. the induced orthorhomicity). As described in the main text, the proportionality constant is given by the elastoresistivity coefficient $2m_{66}$, and the linearity is verified for the given range of strains over the entire temperature range of the experiment. The microscopic physics that determines this proportionality constant is identical in both the spontaneous orthorhombic state (condensed nematic order) and the strained tetragonal state (strain induced anisotropy due to the bilinear coupling of strain to fluctuating nematic order). In other words, our measurements of $2m_{66}$ directly reveal the resistivity anisotropy associated with orthorhombicity in this material. Our conclusions do not depend on the range of induced strains being equal to that which develops spontaneously in the orthorhombic state, though as explained above they are in fact comparable.


\begin{thebibliography}{10}
\bibitem{constant} It remains to be seen how the proportionality constant $c$ quantitatively depends on disorder in the hot-spots model, which requires a quantitative solution of the Boltzmann transport equation in or beyond the dirty limit.
\end{thebibliography}
\end{document}